\documentclass[reprint, amsmath,amssymb,aps,prd]{revtex4-1}
\usepackage{subfigure}
\usepackage{graphicx}% Include figure files
\usepackage{dcolumn}% Align table columns on decimal point

\usepackage{bm}% bold math
\usepackage{xcolor}
\usepackage[colorlinks,urlcolor=aa,linkcolor=blue,anchorcolor=aa,citecolor=aa]{hyperref}
\usepackage[mathlines]{lineno}% Enable numbering of text and display math
\usepackage{multirow}
\usepackage{enumerate}
\usepackage{amsmath}

\usepackage{lineno}
\usepackage{orcidlink}

\usepackage{mathtools}
\usepackage{overpic}
\usepackage{epsfig}
\usepackage{rotating}
\usepackage{longtable}
\usepackage{color}
\usepackage{siunitx}
\usepackage{diagbox}

\definecolor{aa}{RGB}{0,0,139}

\newcommand{\gev}{\rm GeV}
\newcommand{\gevc}{{\rm GeV}/c}
\newcommand{\gevcs}{{\rm GeV}/c^2}
\newcommand{\mev}{\rm MeV}
\newcommand{\mevcs}{{\rm MeV}/c^2}

\newcommand{\g}{\gamma}
\newcommand{\sqs}{\sqrt{s}}

\newcommand{\KP}{K^+}
\newcommand{\KM}{K^-}
\newcommand{\Kstar}{\bar{K}^{*}(892)^{0}}
\newcommand{\KK}{K^{+}K^{-}}
\newcommand{\pipi}{\pi^{+}\pi^{-}}
\newcommand{\Kpi}{K^{+}\pi^{-}}
\newcommand{\Jpc}{J^{PC}}
\newcommand{\Ds}{D^{+}_{s}}
\newcommand{\Dsone}{D_{s1}^{-}(2536)}
\newcommand{\Dstar}{\bar{D}^{*0}}
\newcommand{\DstoKKpi}{D_{s}^{+}\to K^{+}K^{-}\pi^{+}}
\newcommand{\DstoKsK}{D_{s}^{+} \to K_{S}^{0}K^{+}}

\newcommand{\Ks}{K^{0}_{S}}
\newcommand{\KKpi}{K^{+}K^{-}\pi^{+}}
\newcommand{\KsK}{K_{S}^{0}K^{+}}
\newcommand{\gDsK}{\gamma D_{s}^{+} K^{-}}

\newcommand{\cm}{\si{\centi\metre}}
\newcommand{\ns}{\si{\nano\second}}
\newcommand{\ps}{\si{\pico\second}}

\newcommand{\piz}{\pi^0}

\newcommand{\EE}{e^+e^-}

\newcommand{\gisr}{\gamma_{\rm ISR}}

\newcommand{\reduline}{\bgroup\markoverwith
{\textcolor{red}{\rule[0.5ex]{2pt}{0.4pt}}}\ULon}

\newcommand{\beq}{\begin{equation}}
\newcommand{\eeq}{\end{equation}}
\newcommand{\beqar}{\begin{eqnarray}}
\newcommand{\eeqar}{\end{eqnarray}}
\newcommand{\bitm}{\begin{itemize}}
\newcommand{\eitm}{\end{itemize}}

\def\NIMA{Nucl. Instrum. Methods A}
\def\NPA{Nucl. Phys. A}

\def\PLB{Phys. Lett. B}
\def\PRL{Phys. Rev. Lett.}
\def\PRD{Phys. Rev. D}
\def\PRP{Phys. Rept.}
\def\CPC{Chin. Phys. C}

\def\EPJC{Eur. Phys. J. C}

\def\PTEP{ Prog. Theor. Exp. Phys. }
\def\RMP{Rev. Mod. Phys. }
\def\MPLA{Mod. Phys. Lett. A }

\begin{document}

%\preprint{APS/123-QED}

\title{{\bf \boldmath Search for a $1^{-+}$ molecular state via $\EE \to \g \Ds \Dsone +c.c.$}}

%% Saved at => 2024-08-06
\author{M.~Ablikim$^{1}$, M.~N.~Achasov$^{4,c}$, P.~Adlarson$^{76}$, O.~Afedulidis$^{3}$, X.~C.~Ai$^{81}$, R.~Aliberti$^{35}$, A.~Amoroso$^{75A,75C}$, Y.~Bai$^{57}$, O.~Bakina$^{36}$, I.~Balossino$^{29A}$
% ...
\\
\vspace{0.2cm}
(BESIII Collaboration)\\
\vspace{0.2cm} {\it
$^{1}$ Institute of High Energy Physics, Beijing 100049, People's Republic of China\\
% ...
$^{3}$ Bochum  Ruhr-University, D-44780 Bochum, Germany\\
$^{4}$ Budker Institute of Nuclear Physics SB RAS (BINP), Novosibirsk 630090, Russia\\
% ...
$^{29}$ INFN Sezione di Ferrara, (A)INFN Sezione di Ferrara, I-44122, Ferrara, Italy; (B)University of Ferrara,  I-44122, Ferrara, Italy\\
% ...
$^{35}$ Johannes Gutenberg University of Mainz, Johann-Joachim-Becher-Weg 45, D-55099 Mainz, Germany\\
$^{36}$ Joint Institute for Nuclear Research, 141980 Dubna, Moscow region, Russia\\
% ...
$^{57}$ Southeast University, Nanjing 211100, People's Republic of China\\
% ...
$^{75}$ University of Turin and INFN, (A)University of Turin, I-10125, Turin, Italy; (B)University of Eastern Piedmont, I-15121, Alessandria, Italy; (C)INFN, I-10125, Turin, Italy\\
$^{76}$ Uppsala University, Box 516, SE-75120 Uppsala, Sweden\\
% ...
$^{81}$ Zhengzhou University, Zhengzhou 450001, People's Republic of China\\
\vspace{0.2cm}
% ...
$^{c}$ Also at the Novosibirsk State University, Novosibirsk, 630090, Russia\\
% ...
}}
%% ends here %%

\vspace{0.4cm}
\date{\today}
\begin{abstract}

%The molecular state $X$, with quantum numbers $\Jpc=1^{-+}$, is searched for via the process $\EE \to \g \Ds \Dsone +c.c.$ for the first time across center-of-mass energies from 4.612 to 4.951~$\gev$, using $5.8~\mathrm{fb^{-1}}$ data samples collected with the BESIII detector operating at the BEPCII collider.
We search, for the first time, for an exotic molecular state with quantum numbers $\Jpc=1^{-+}$, called $X$, via the process $\EE \to \g \Ds \Dsone +c.c.$ using data samples corresponding to a luminosity of $5.8~\mathrm{fb^{-1}}$ across center-of-mass energies from 4.612 to 4.951~$\gev$, collected with the BESIII detector operating at the BEPCII collider.
No statistically significant signal is observed. The upper limits on the product of cross-section and branching fraction $\sigma({\EE \to \g X}) \cdot \mathcal{B}(X \to \Ds \Dsone +c.c.)$ at 90\% confidence level are reported for each energy point, assuming the $X$ mass to be 4.503~$\gevcs$ and the width 25, 50, 75, and 100~$\mev$, respectively.
\end{abstract}

\pacs{Valid PACS appear here}
\maketitle

%\tableofcontents
\section{Introduction}
\label{sec:introduction}
Since their discovery two decades ago \cite{charmonium_review}, the charmonium-like states, known as XYZ, have enormously broadened our understanding of the hadronic mass spectrum. Unlike the conventional charmonium states, which are composed of charm quark anti-quark pairs ($c\bar{c}$), the XYZ states are believed to present a more complex internal structure, including e.g. tetraquark, molecule, or hybrid. Therefore, they provide additional information which goes beyond the traditional $c\bar{c}$ systems. The investigation of their spectrum, quantum numbers, production rate, and decays can shed light on the mechanisms of the strong interaction.

Since the charmonium-like states have the same quantum numbers as the charmonium states, they are difficult to be distinguished.
For instance, the $X(3872)$ was first observed in the $B^{\pm} \to K^{\pm}\pipi J/\psi$ decay by Belle~\cite{X3872_2003} in 2003 and subsequently confirmed by several other experiments~ \cite{X3872_CDF,X3872_D0,X3872_BaBar}. While $X(3872)$ is believed as the first exotic charmonium-like particle~\cite{X_mole1,X_mole2,X_mixed1,X_conven,X_tetra,X_mixed2,X_mixed3}, there remains a long-standing debate about whether it could
instead be the conventional $\chi_{c1}(2P)$ state~\cite{X_conven}.
Similarly, the vector state $Y(4230)$, which was observed and subsequently confirmed in its decay to $\pipi J/\psi$~\cite{Y4230_BaBar, Y4230_CLEO, Y4230_Belle};
there are ongoing arguments suggesting that it could be the charmonium $\psi(4S)$ state~\cite{Y4230_cc1,Y4230_cc2,Y4230_cc3,Y4230_cc4}, or an exotic state~\cite{Y4230_mole1,Y4230_mole2,Y4230_mole3,Y4230_hybrid1,Y4230_hybrid2}.

One possible way to bypass the aforementioned difficulty is to search for states that are ``more exotic than the other exotic states''. The $Z^\pm_c(3900)$~\cite{BESIII:2013ris}
 is one kind of these states since it is charged and contains a $c\bar{c}$ component. In between the light meson states, the $\pi_1(1400)$~\cite{pi1_1400_1,pi1_1400_2,pi1_1400_3,pi1_1400_4,pi1_1400_5}, $\pi_1(1600)$~\cite{pi1_1600_1,pi1_1600_2}, and $\eta_{1}(1855)$~\cite{eta1_1855} are considered exotic states due to their unusual quantum numbers $\Jpc=1^{-+}$,  indicating non-$q\bar{q}$ quark components.
Exotic states with unusual quantum numbers have been found only in the light hadron spectrum, and up till now, no similar states have been discovered in the charmonium energy region, even if they are allowed by QCD.
Ref.~\cite{WQ} discusses the possibility that heavy-light meson states, such as $\bar{D} D_1(2420)$ and its charge conjugate ($c.c.$), can couple to states with exotic quantum numbers $\Jpc=1^{-+}$ in $S$-wave. Similarly, this can also be extended to $D$-mesons with a strange quark, suggesting potential heavier exotic $1^{-+}$ states near the $\Ds \Dsone$ threshold. Throughout this paper without specification, the charge conjugated mode is implied. 

In this paper, the process $\EE \to \g \Ds \Dsone$ has been used to search for a molecular $1^{-+}$ state $X$, formed by $\Ds \Dsone$. This has been done using twelve data samples with center-of-mass (c.m.) energies $(\sqs)$ ranging from 4.612 to 4.951~$\gev$ \cite{XYZ_data}, corresponding to an integrated luminosity of $5.8~\mathrm{fb^{-1}}$. The values of c.m.\ energy and luminosity are listed in Table \ref{table:up_final}.      

\section{BESIII Detector and Monte Carlo simulation}
\label{detector}

The BESIII detector~\cite{Ablikim:2009aa} records symmetric $e^+e^-$ collisions 
provided by the BEPCII storage ring~\cite{BESIII:2020nme}
in the c.m. energy range from 1.84 to 4.95~$\gev$,
with a peak luminosity of $1.1 \times 10^{33}\;\text{cm}^{-2}\text{s}^{-1}$ 
achieved at $\sqrt{s} = 3.773\;\text{GeV}$. BESIII has collected large data samples in this energy region~\cite{BESIII:2020nme}. The cylindrical core of the BESIII detector covers 93\% of the full solid angle and consists of 
a helium-based multilayer drift chamber~(MDC), a plastic scintillator time-of-flight system~(TOF), and a CsI(Tl) 
electromagnetic calorimeter~(EMC), which are all enclosed in a superconducting solenoid magnet providing a 1.0~T 
magnetic field. The solenoid is supported by an octagonal flux-return yoke with resistive plate counter muon 
identification modules interleaved with steel. %The acceptance of charged particles and photons is 93\% of the $4\pi$ solid angle.
The charged-particle momentum resolution at $1~\gevc$ is $0.5\%$, and the $dE/dx$ resolution is 6\% for 
electrons from Bhabha scattering. The EMC measures photon energies with a resolution of 2.5\% (5\%) at $1~\gev$ in 
the barrel (end cap) region. The time resolution in the TOF barrel region is $68~\ps$, while that in the end cap region is $110~\ps$. The end cap TOF system was upgraded in 2015 using multi-gap resistive plate chamber technology, providing a 
time resolution of $60~\ps$, which benefits the total amount of the data used in this analysis~\cite{etof}. 

Simulated data samples produced with a {\sc
geant4}-based~\cite{geant4} Monte Carlo (MC) package, which includes the geometric description of the BESIII detector~\cite{BESIII:detector_descrip} and the detector response, are used to determine the detection efficiencies and estimate backgrounds. The simulation models the beam energy spread and ISR in 
the $\EE$ annihilation with the {\sc kkmc} generator~\cite{ref:kkmc}. The inclusive MC sample includes the production of open charm
processes, the ISR production of vector charmonium(-like) states, and the continuum processes incorporated in {\sc
kkmc}~\cite{ref:kkmc}. All the known particle decays are modeled with {\sc evtgen}~\cite{ref:evtgen} using the branching fractions either taken from 
the Particle Data Group (PDG)~\cite{PDG} or estimated with {\sc lundcharm}~\cite{ref:lundcharm}. 
The final state radiation~(FSR) from charged final state particles is incorporated using the {\sc photos} package~\cite{photos}. 

We generate 200,000 signal MC events of $\EE \to \g X$ and $X \to \Ds \Dsone$ at each c.m.\ energy and each hypothetical $X$'s width with uniform phase space (PHSP) distribution. The $X$'s mass ($M_{X}$) is set to the $\Ds\Dsone$ mass threshold 4.503~$\gevcs$, based on the molecule assumption that its mass should be very close to, perhaps a few MeV lower than, this threshold~\cite{WQ}; its width ($\Gamma_{X}$) is set to 25, 50, 75, and 100~$\mev$, reflecting the widths of the other charmonium-like states. We generate the $\DstoKKpi$ decay based on the amplitude analysis results from Refs.~\cite{ref:Ds_Daliz_CLEO,ref:Ds_Daliz_BES,ref:Ds_Daliz_BaBar}, and the $\DstoKsK$ and $\Ks \to \pipi$ decays according to the PHSP distribution. 
We generate $\Dsone \to \Dstar \KM$ process via VVS-PWAVE model~\cite{ref:evtgen}, with inclusive decays of $\Dstar$ following the world-averaged branching fractions~\cite{PDG}, where $\Dstar$ decays 64.7\% into $\bar{D}^{0}\piz$ and 35.3\% into $\bar{D}^{0}\g$. The cross section line shape is assumed to be proportional to $E_{\g}^{3}/s$~\cite{ref:Yangy}, where $E_\g$ is the energy of the radiative photon. This information is used to obtain the radiative correction factor and detection efficiency. 

\section{Primary event selection}
\label{sec:selection}
We employ a partial reconstruction method for the signal process $\EE \to \g \Ds \Dsone$, $\Dsone \to \Dstar \KM$ to achieve higher efficiency. This method involves reconstructing the $\g$, $\Ds$, and a bachelor $\KM$ from the $\Dsone$ decay. The $\Dsone$ and its daughter particle $\Dstar$ are searched for in the recoiling mass spectrum of the $\g\Ds$ and $\g \Ds \KM$ candidates, respectively. The $\Ds$ meson is reconstructed via $\KKpi$ or $\Ks \KP$ with $\Ks \to \pipi$. 

Photon candidates are identified using showers in the EMC. The deposited energy of a shower must be greater than $25~\mev$ 
in the barrel region ($|\cos \theta|< 0.80$) or greater than $50~\mev$ in the end cap region ($0.86 <|\cos \theta|< 0.92$). Here $\theta$ is the polar angle with respect to the $z$-axis, the symmetry axis of the MDC.
To exclude showers that originate from charged tracks, the angle subtended by the EMC shower and the position of the 
closest charged track at the EMC must be greater than 10 degrees as measured from the interaction point (IP). To suppress the 
electronic noise and the showers unrelated to the event, the difference between the EMC time and the event start time is required to be within $[0,~700]~\ns$. The number of photons per event is required to be at least one.

A charged track is reconstructed from the hits in the MDC. We require that each charged track not associated with the $\Ks$ must 
satisfy $|\cos \theta|<0.93$, and the distance of the closest approach to the IP within $10~\cm$ along the $z$-axis ($V_{z}$), and less than
$1~\cm$ in the transverse plane. 
Particle identification (PID) for charged tracks combines measurements of the energy deposited in the MDC~(d$E$/d$x$) and the time-of-flight in the TOF to form likelihoods $\mathcal{L}(h)~(h=p,K,\pi)$ for each hadron $h$ hypothesis. Tracks are identified as $K^{\pm}$ ($\pi^{\pm}$) by comparing the likelihoods for the kaon and pion hypotheses, requiring $\mathcal{L}(K)>\mathcal{L}(\pi)$ and $\mathcal{L}(K)>0$ ($\mathcal{L}(\pi)>\mathcal{L}(K)$ and $\mathcal{L}(\pi)>0$).

Each $\Ks$ candidate is reconstructed from two oppositely charged tracks that satisfy $|V_{z}| < 20$ cm and $|\cos \theta|<0.93$. The two charged tracks are assigned as $\pipi$ without imposing any PID criteria. 
They are constrained to originate from a common vertex and are required to have an invariant mass
within $|M_{\pipi} - m_{\Ks}|< 9~\mevcs$, where
$m_{\Ks}$ is the $\Ks$ nominal mass~\cite{PDG}. The
decay length of the $\Ks$ candidate is required to be greater than
twice the vertex resolution away from the IP.

\begin{figure*}[!htbp]
\centering   
	\begin{overpic}[angle=0,width=0.45\textwidth]{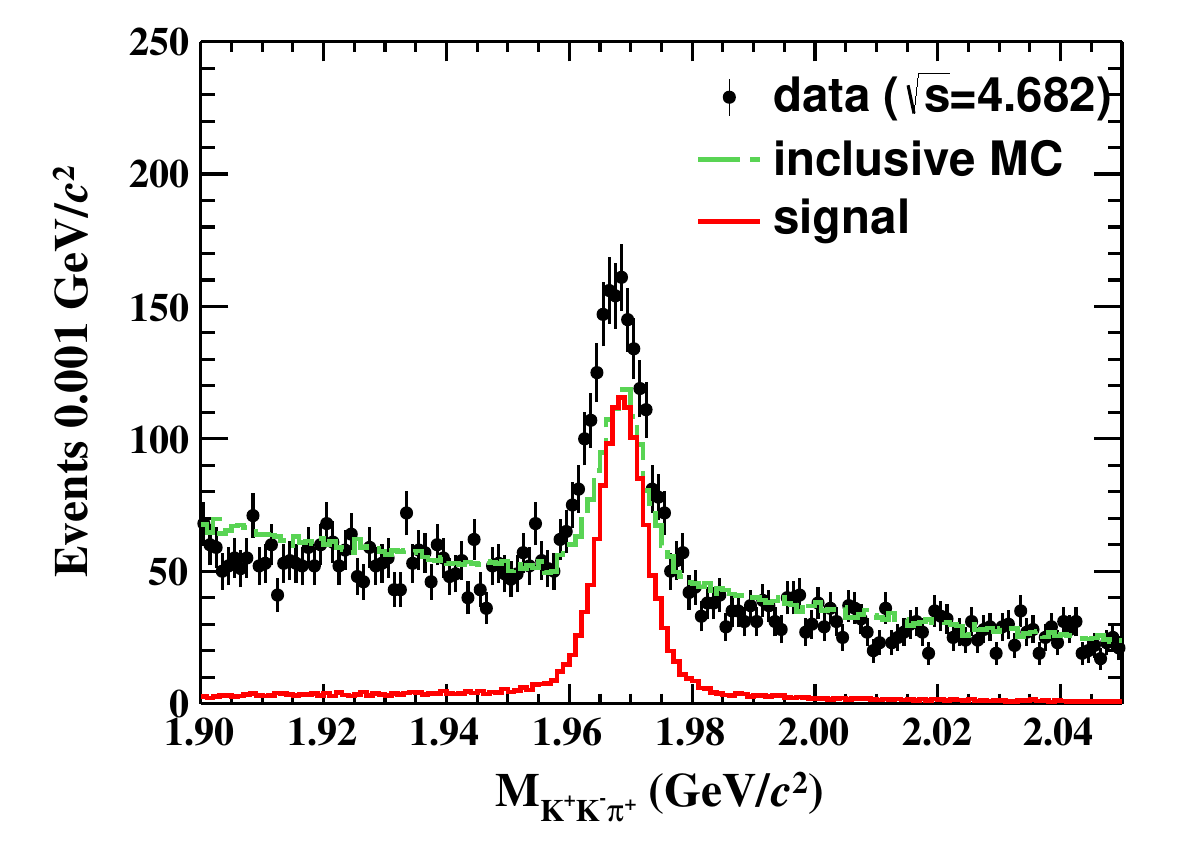}
	  \put(20,60){$(a)$}
	\end{overpic}
	\begin{overpic}[angle=0,width=0.45\textwidth]{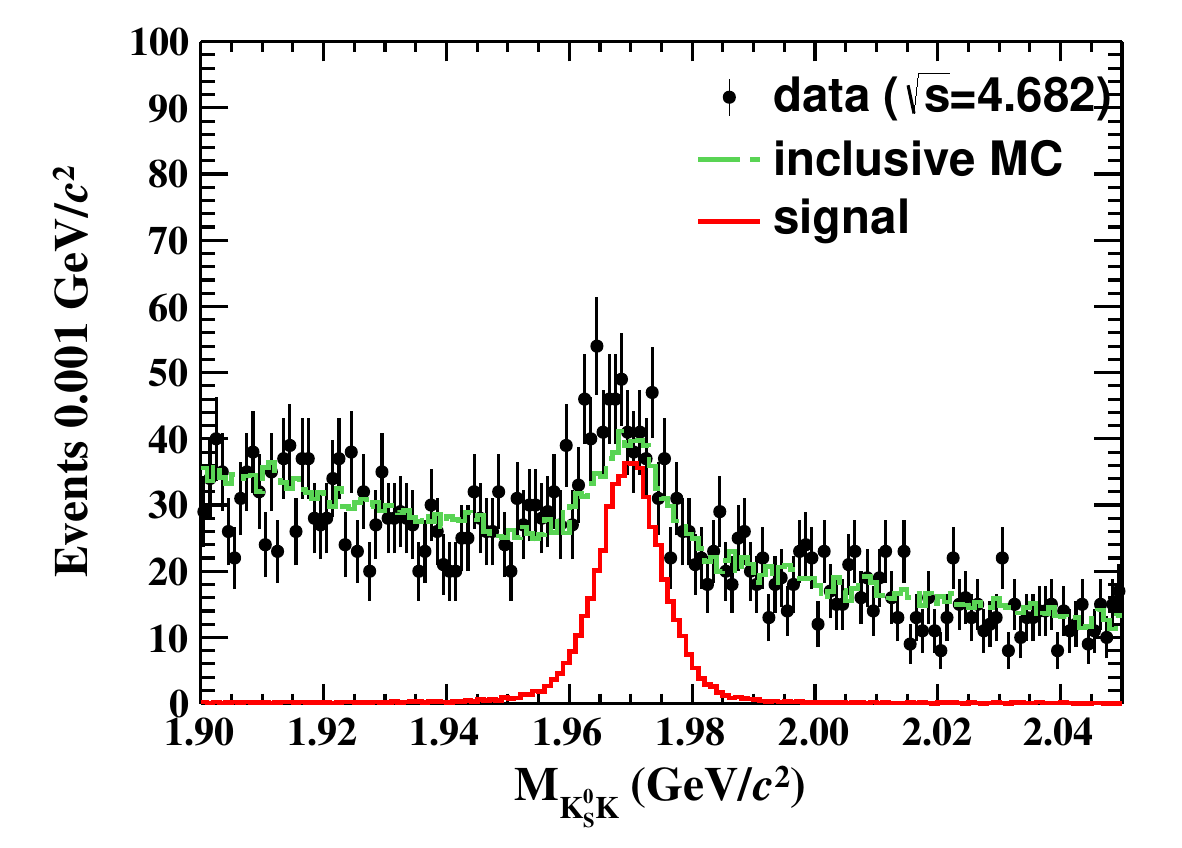}
	 \put(20,60){$(b)$}
	\end{overpic}
\caption{Invariant mass distributions of  $M_{\KKpi}$ $(a)$ and $M_{\KsK}$ $(b)$ at $\sqrt{s}=4.682~\gev$. Dots with error bars represent data, the green dashed lines represent inclusive MC, red solid lines indicate the signal. The number of events in the inclusive MC simulation is normalized to match the data, while the signal is normalized arbitrarily.}
\label{Fig:mass_Ds}
\end{figure*}

The selected $K^{\pm}$, $\pi^{\pm}$, and $\Ks$ candidates in an event are combined to reconstruct $\DstoKKpi$ or $\DstoKsK$, denoted as the $KK\pi$ or $\Ks K$ modes, respectively. At least one $\KM$ with opposite charge to the $\Ds$ candidate is required. Only the decays containing the intermediate states $\phi$ or $\Kstar$ in the $KK\pi$ mode are used to select the $\Ds$ candidates. The invariant mass of $\KK$ ($\Kpi$) is required to satisfy $|M_{\KK}-m_{\phi}|< 9~\mevcs$ 
($|M_{\Kpi}-m_{\Kstar}|< 66~\mevcs$). %Since $\phi \to \KK$ and $\Kstar \to \Kpi$ are through $P$-wave \cite{ref:evtgen}, the helicity angle can be utilized to suppress backgrounds. 
The helicity angle of $\KP$ in the $\KK$ $(\Kpi)$ helicity frame is required to satisfy $|\cos \theta|>0.36$ ($|\cos \theta|>0.45$) to improve the significance of the $D_s$ meson. As an example, Fig.~\ref{Fig:mass_Ds} shows the invariant mass distributions of $\KKpi$ ($M_{\KKpi}$) and $\KsK$ ($M_{\KsK}$) at $\sqrt{s} = 4.682$ GeV. The masses $M_{\KKpi}$ and $M_{\KsK}$ must satisfy $|M_{\KKpi}-m_{\Ds}|< 15~\mevcs$ and $|M_{\KsK}-m_{\Ds}|< 19~\mevcs$, respectively.
Here and hereafter, $m_{y}$ $(y=\phi, \Kstar, \Ds)$ denotes the respective nominal masses~\cite{PDG}. 
Since the resolutions and background levels are dependent on the c.m.\ energy, we categorize the data samples into three sets based on $\sqs$ to optimize the event selection criteria:
\begin{itemize}
     \item Set I: 4.612, 4.628, 4.641, 4.661, 4.682 GeV;
     \item Set II: 4.699, 4.740, 4.750, 4.781 GeV; 
     \item Set III: 4.843, 4.918, 4.951 GeV.
\end{itemize}
Within a set, the resolutions and backgrounds are assumed to be similar for each energy point. To improve the resolution, the modified recoiling mass of $\gDsK$ is defined as $RM^{'}_{\Dstar} \equiv RM_{\gDsK}+M_{\KKpi}-m_{\Ds}$, with $RM_{\gDsK}=\sqrt{(p_{c.m.}-p_{\g}-p_{\Ds}-p_{\KM})^2}$, in which $p_{c.m.}$, $p_{\g}$, $p_{\Ds}$, and $p_{\KM}$ are the four-momenta of the initial $\EE$ system, the radiative photon, $\Ds$, and $\KM$, respectively. This definition is specified for the $KK\pi$ mode, and a similar definition is applied for the $\Ks K$ mode. The interval range requirements of $RM^{'}_{\Dstar}$ in the three sets are presented in Table~\ref{table:RM_gDsK_chi_cut}. These requirements are based on the Punzi method~\cite{Punzi}, optimizing the figure-of-merit (FOM) $\epsilon/(\alpha/2+\sqrt{B})$, where $\alpha=3$ indicates the expected significance, $\epsilon$ is the selection efficiency, and $B$ is the number of background events from inclusive MC, as will be discussed in the next section. All possible combinations are retained for later analysis.

\begin{figure*}[!htbp]
\centering   
	\begin{overpic}[angle=0,width=0.45\textwidth]{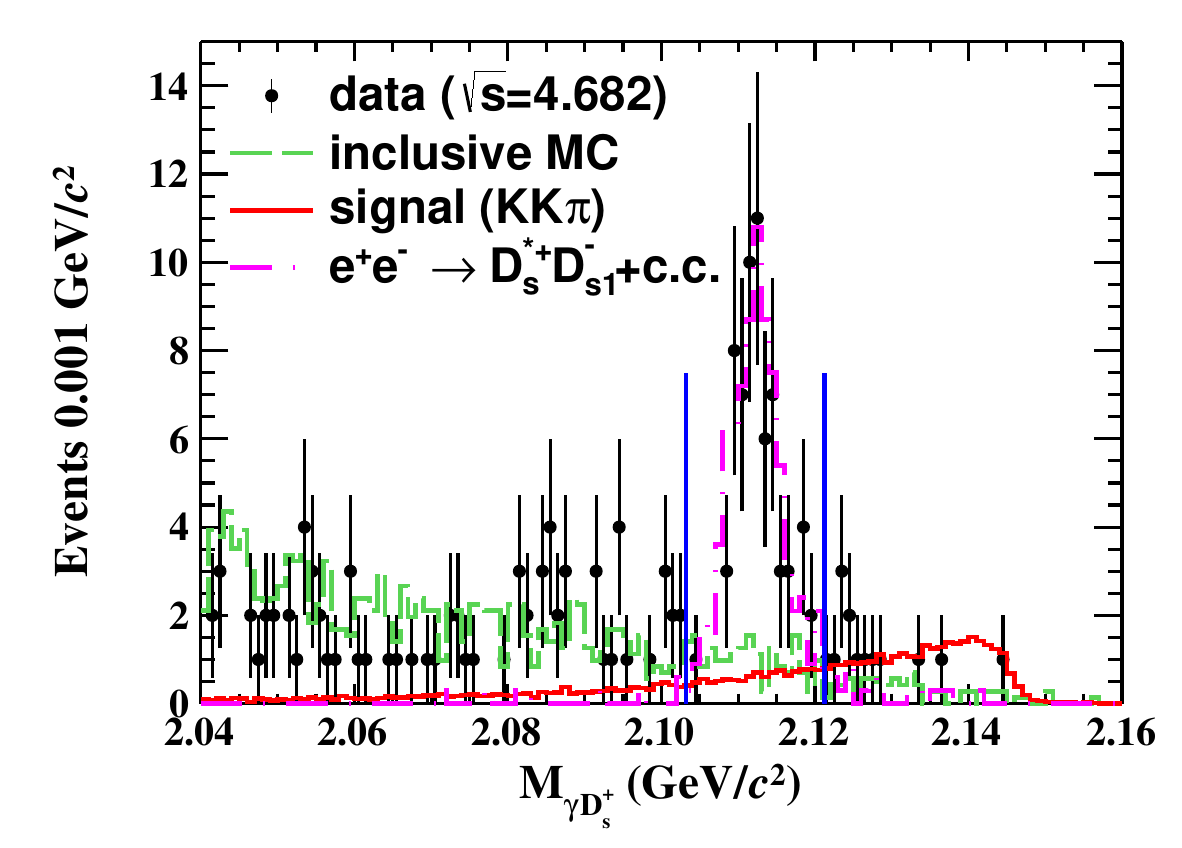}
	  \put(80,60){$(a)$}
	\end{overpic}
	\begin{overpic}[angle=0,width=0.45\textwidth]{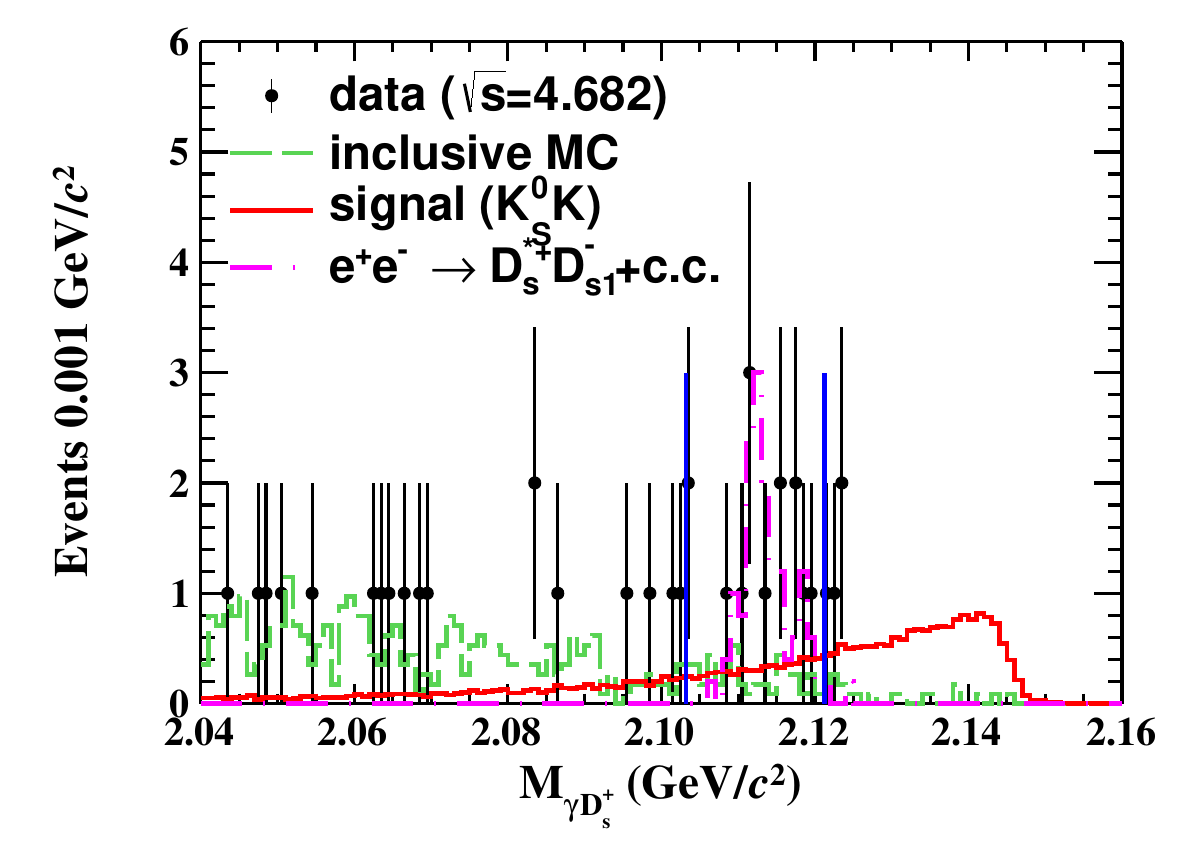}
	 \put(80,60){$(b)$}
	\end{overpic}
\caption{Distributions of $M_{\g\Ds}$ at $\sqrt{s}=4.682~\gev$ for the $KK\pi$ mode $(a)$ and the $\Ks K$ mode $(b)$ after the primary event selection. Dots with error bars indicate data, the green dashed lines indicate inclusive MC, red solid lines indicate signal, the pink dashed line indicates the peaking background from $\EE \to D_{s}^{*+} \Dsone, D_{s}^{*+} \to \g\Ds$, and blue solid lines indicate the mass range requirement for $M_{\g\Ds}$.}
\label{Fig:mass_gDs}
\end{figure*}

\begin{figure*}[!htbp]
\centering   
	\begin{overpic}[angle=0,width=0.45\textwidth]{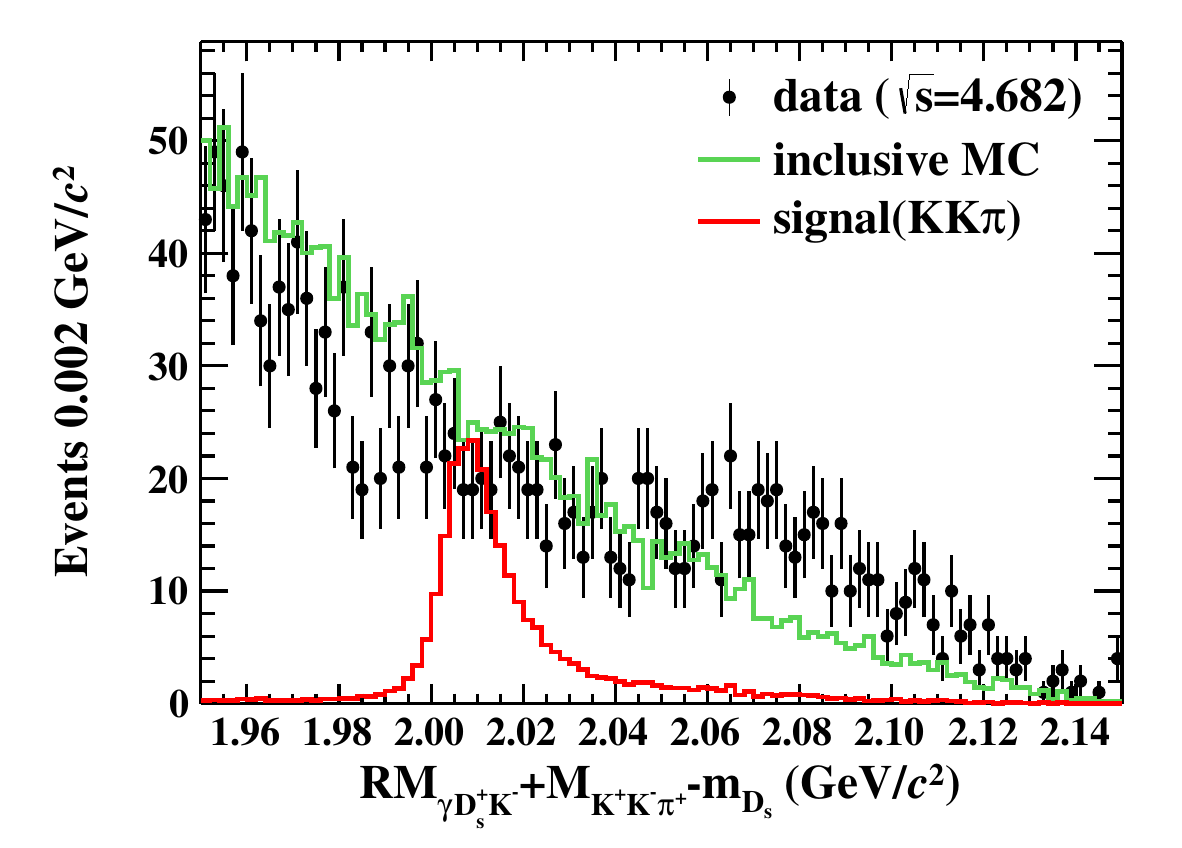}
	  \put(30,60){$(a)$}
	\end{overpic}
	\begin{overpic}[angle=0,width=0.45\textwidth]{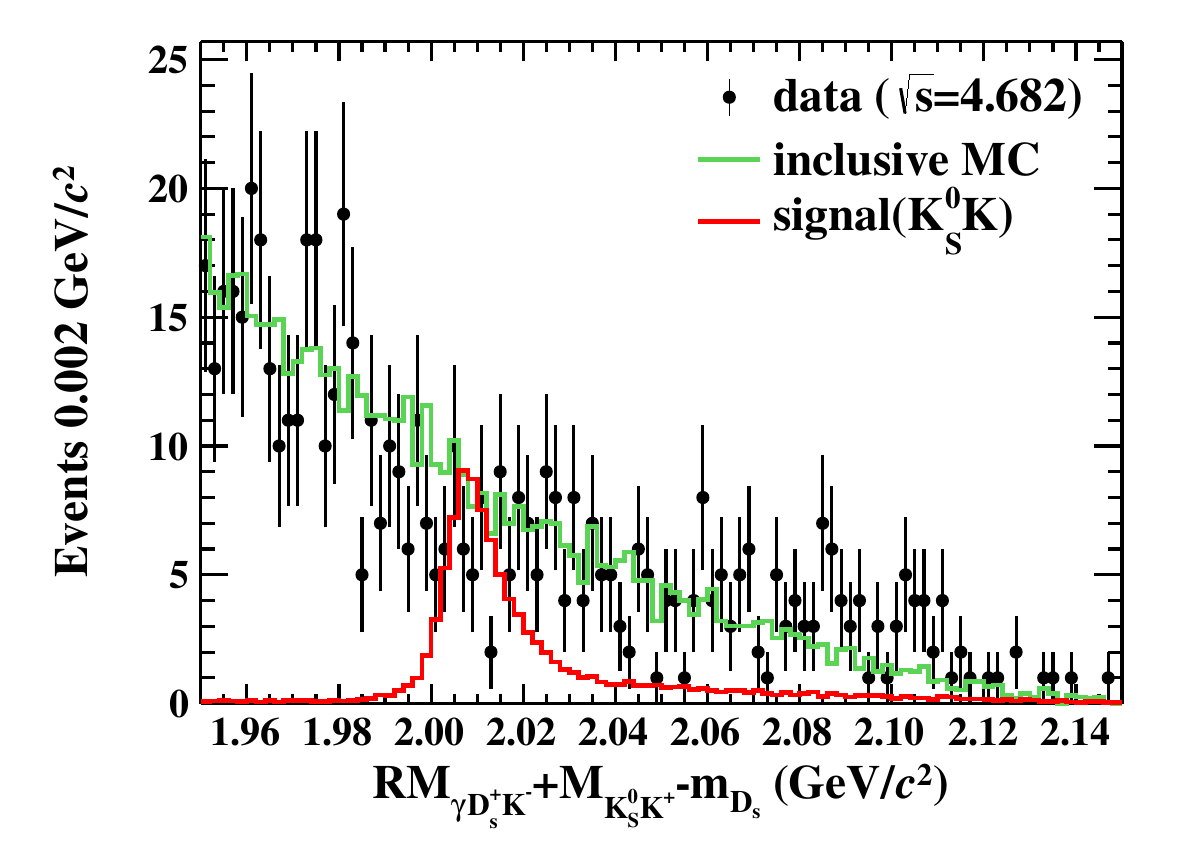}
	 \put(30,60){$(b)$}
	\end{overpic}
\caption{Distributions of $RM^{'}_{\Dstar}$ at $\sqrt{s}=4.682~\gev$ for the $KK\pi$ mode $(a)$ and the $\Ks K$ mode $(b)$ after the requirement of $M_{\g\Ds}$. Dots with error bars indicate data, the green dashed lines indicate inclusive MC, red solid lines indicate signal. The number of events in the inclusive MC simulation is normalized to match the data, while the signal is normalized arbitrarily.}
\label{Fig:RM_gDsK}
\end{figure*}
 
 \section{Background analysis and suppression}
 \label{sec:bg}
 Based on~\cite{ref:gDstar}, the $\EE \to D_{s}^{*+} \Dsone, D_{s}^{*+} \to \g\Ds$ events constitute a peaking background and significantly contaminate the signal, since the final state is similar to the signal one. We require the invariant mass of $\g\Ds$ ($M_{\g\Ds}$) to satisfy $|M_{\g\Ds}-m_{D_{s}^{*+}}|> 9~\mevcs$ to suppress this kind of background, as shown in Fig.~\ref{Fig:mass_gDs}, where $m_{D_{s}^{*+}}$ is the nominal mass of $D_{s}^{*+}$~\cite{PDG} and the data sample at $\sqrt{s} = 4.682$ GeV is presented as an example. It should be noted that suppressing this peaking background leads to a reduced efficiency at the energy values $\sqrt{s}=4.661$ and $4.682$, as a result of the relatively larger overlap between the $M_{\g\Ds}$ distributions of the signal and background. After imposing all the above event selection criteria, the distributions of $RM^{'}_{\Dstar}$ are shown in Fig.~\ref{Fig:RM_gDsK}. 
 We generate the exclusive MC samples of $\EE \to D^{*+}_{s} \Dsone$ with $D^{*+}_{s} \to \g \Ds$ and $\EE \to \Ds \Dsone$ processes to estimate the background contamination. Their production cross-section line shapes and decay models are taken from the two BESIII measurements~\cite{ref:gDstar,ref:whp}.
 The number of events for $\EE \to D_{s}^{*+} \Dsone, D_{s}^{*+} \to \g\Ds$ at the 4.682~$\gev$ energy is estimated to be 2.5 in the $KK\pi$ mode and 0.5 in the $\Ks K$ mode by simulation. In addition, the backgrounds from $\EE \to (\gisr)\Ds\Dsone$ are determined to be 1.2 events in the $KK\pi$ and 0.3 in the $\Ks K$ mode at 4.682~$\gev$. Therefore, we ignore these two kinds of background in the cross-section measurements, and only consider their impact in the estimation of the systematic uncertainties.
 
To improve the resolution and further suppress the background, we perform a two-constraint (2C) kinematic fit  to all the selected candidates, constraining $M_{\KKpi/\KsK}$ to $m_{\Ds}$ and $RM_{\gDsK}$ to $m_{\Dstar}$. Here, $m_{\Dstar}$ is the nominal mass of $\Dstar$~\cite{PDG}. If there are more than one combination in the event, the candidate with the lowest $\chi^{2}$ value is chosen. To further suppress the backgrounds, we employed the Punzi method~\cite{Punzi} to optimize the $\chi^{2}$ requirements, as shown in Table~\ref{table:RM_gDsK_chi_cut}. 

\begin{table*}[!htbp]
\setlength{\abovecaptionskip}{1pt} 
\setlength{\belowcaptionskip}{4pt}
\caption{The values used to determine $\sigma(\EE \to \g X) \cdot \mathcal{B}(X \to \Ds \Dsone)$. Only the statistical uncertainty is presented, and the mass and width of the $X$ candidate are set to be $4.503$~$\gevcs$ and $0.05$~$\gev$, respectively. $\mathcal{L}_{int}$ is the integral luminosity, $(1+\delta)$ is the radiative correction factor, $\frac{1}{|1-\Pi|^2}$ is the VP factor, $\epsilon_{KK\pi}$ ($\epsilon_{K_{S}^{0}K}$) is the selection efficiency of the $KK\pi$ ($K_{S}^{0}K$) mode, $N_{KK\pi}$ is the number of signal events for the $KK\pi$ mode obtained by fitting, $S$ is the statistical significance of the signal, $N_{KK\pi}^{UL}$ is the upper limit of the number of signals at 90\% C.L., and $\sigma^{UL} \cdot \mathcal{B}$ is the upper limit of the product of the cross-section and the branching fraction at 90\% C.L.}
\centering
%\footnotesize 
\setlength{\tabcolsep}{5pt}
\renewcommand{\arraystretch}{1.5}
\begin{tabular}{c c c c c c c c c c}
\hline \hline
$\sqs~(\gev)$  &$\mathcal{L}_{int}$ $({\rm pb^{-1}})$ &$(1+\delta)$ &$\frac{1}{{|1-\prod|}^2}$  &$\epsilon_{KK\pi} (\%)$  &$\epsilon_{K_{S}^{0}K} (\%)$ &$N_{KK\pi}$ & $S$  &$N_{KK\pi}^{UL}$   &$\sigma^{UL} \cdot \mathcal{B}$ (pb)  \\
\hline
4.612  &103.7  &0.689  &1.055  &4.47   &8.04   &$0.0^{+0.4}_{-0.0}$  &0.1$\sigma$  &1.7  &26.8  \\
4.628  &521.5  &0.734  &1.054  &4.27   &7.66   &$0.6^{+1.3}_{-0.7}$  &0.1$\sigma$  &2.8  &8.6  \\
4.641  &551.7  &0.751  &1.054  &4.03   &7.36   &$1.4^{+1.8}_{-1.0}$  &1.8$\sigma$  &5.4  &16.0  \\
4.661  &529.4  &0.773  &1.054  &2.60   &4.61   &$0.0^{+0.6}_{-0.0}$  &0.3$\sigma$  &2.8  &13.1  \\
4.682  &1667.4  &0.785  &1.054  &3.58  &6.12   &$7.0^{+3.5}_{-2.8}$  &3.5$\sigma$  &12.5 &13.1  \\
4.699  &535.5  &0.793  &1.054  &4.34   &7.83   &$2.2^{+1.6}_{-1.3}$  &2.1$\sigma$  &5.6  &15.0  \\
4.740  &163.9  &0.808  &1.055  &4.82   &8.63   &$0.0^{+0.4}_{-0.0}$  &1.1$\sigma$  &1.9  &15.0  \\
4.750  &366.6  &0.811  &1.055  &4.75   &8.86   &$0.5^{+2.2}_{-1.3}$  &0.3$\sigma$  &4.3  &15.1  \\
4.781  &511.5  &0.821  &1.055  &4.92   &9.15   &$0.0^{+0.5}_{-0.0}$  &0.9$\sigma$  &2.0  &4.9  \\
4.843  &525.2  &0.835  &1.056  &4.51   &8.64   &$0.0^{+0.4}_{-0.0}$  &0.1$\sigma$  &1.8  &4.6  \\
4.918  &207.8  &0.845  &1.056  &4.79   &8.95   &$2.0^{+1.8}_{-1.2}$  &2.2$\sigma$  &5.1  &30.0  \\
4.951  &159.3  &0.851  &1.056  &4.68   &8.71   &$0.0^{+0.4}_{-0.0}$  &0.1$\sigma$  &1.7  &13.5\\
\hline \hline
\end{tabular}
\label{table:up_final}
\end{table*}

\begin{table*}[!htbp]
\setlength{\abovecaptionskip}{0pt}
\setlength{\belowcaptionskip}{9pt}
\centering
\footnotesize  	% \tiny \scriptsize \footnotesize \small \normalsize \large \Large \LARGE \huge
\setlength{\tabcolsep}{5pt}					% column separation 12pt
\renewcommand{\arraystretch}{1.5} 	% row space
\caption{The requirements on the modified recoiling mass $RM^{'}_{\Dstar}$ and on the $\chi^2$  of the 2C kinematic fit for each mode and energy set.}
\begin{tabular}{c c c c}
\hline \hline
Variables\slash Sets          &I  & II   &III \\
\hline 
$RM^{'}_{\Dstar}$ for $KK\pi$ $(\mevcs)$     &$(1999.9,~2034.9)$ &$(1999.9,~ 2029.9)$ &$(1993.9,~2041.9)$\\
$RM^{'}_{\Dstar}$ for $\Ks K$ $(\mevcs)$    &$(2001.9,~2021.9)$
&$(1997.9,~2033.9)$ &$(1994.9,~2041.9)$\\
\hline 
$\chi^{2}$ in $KK\pi$       &$<12.4$   &$<16.4$    &$<8.4$\\
$\chi^{2}$ in $\Ks K$       &$<20.4$   &$<23.2$    &$<12.4$\\
\hline \hline
\end{tabular}
\label{table:RM_gDsK_chi_cut}
\end{table*}

\begin{figure*}[!htbp]
\centering  
\begin{overpic}[angle=0,width=0.9\textwidth]{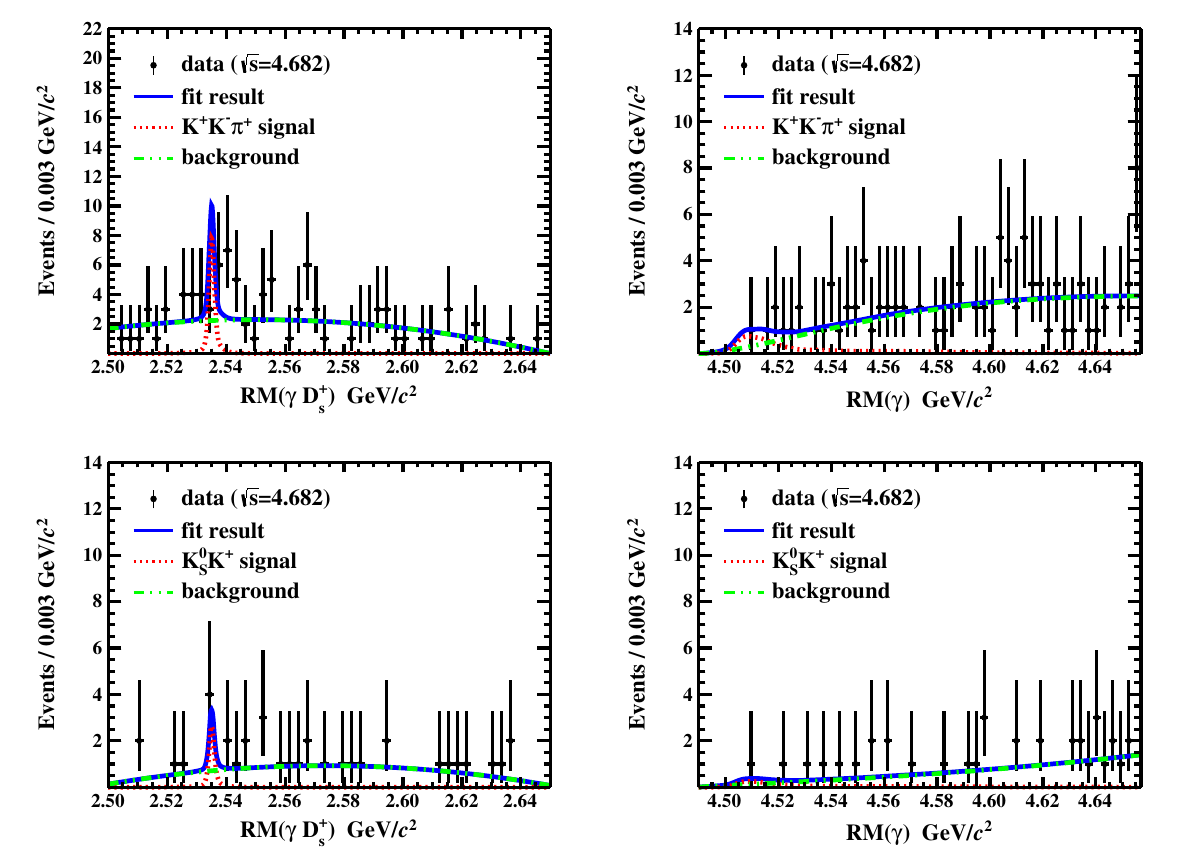}
\put(40,65){$(a)$}
\put(90,65){$(b)$}
\put(40,28){$(c)$}
\put(90,28){$(d)$}
\end{overpic}
\caption{The results of the two-dimensional simultaneous fit at $\sqs=4.682~\gev$, where $(a)$ and $(b)$ are the projected distributions of the $\g\Ds$ recoiling mass and of the $\g$ recoiling mass  for the $KK \pi$ mode, while $(c)$ and $(d)$ are for the $\Ks K$ mode. Dots with error bars indicate data, blue solid curves indicate the best fit results, red dashed lines indicate the $\Dsone$ and $X$ signals, and the green dashed lines indicate the backgrounds.}
\label{Fig:4680_bestfit}
\end{figure*}
 
\section{Signal yields and cross sections} 
\label{sec:sc}
To extract the number of signal events, we perform an unbinned two-dimensional simultaneous fit to the recoiling mass distribution of $\g\Ds$ $(RM_{\g\Ds})$ in the range $(2.50,2.60)$~$\gevcs$ and of $\g$ $(RM_{\g})$ in the range $(4.49,E')$~$\gevcs$, where $E'$ represents $\sqs-0.025~\gev$ (0.025~$\gev$ is the minimum requirement for the photon deposited energy). The line-shapes of the $\Dsone$ and of the $X$ states are described by the simulated MC shapes with  $M_{X}$ set to 4.503~$\gevcs$ and $\Gamma_{X}$ set to 50~$\mev$. For the subsequent study of the systematic uncertainties, additional MC samples are generated with $\Gamma_{X}$ set to $25$, $75$, or $100$~$\mev$. The background is described by a second-order Chebychev function for the $RM_{\g\Ds}$ and a flipped Argus function~\cite{Argus} for the $RM_{\g}$. The signal events in the $KK\pi$ $(N_{KK\pi})$ mode and in the $\Ks K$ $(N_{\Ks K})$ mode are correlated by  
\begin{equation}
N_{\Ks K}=f_{sig} \cdot N_{KK\pi},
\end{equation}
\begin{equation}
f_{sig}= \frac{\epsilon_{\Ks K} \cdot \mathcal{B}(\Ds \to \KsK) \cdot \mathcal{B}(\Ks \to \pipi) }{\epsilon_{KK\pi} \cdot \mathcal{B}(\Ds \to \KKpi)}\ ,
\end{equation}
where $\epsilon_{\Ks K}$ and $\epsilon_{KK\pi}$ are the efficiencies of event selection in the $\Ks K$ and $KK\pi$ modes, while $\mathcal{B}(\Ds \to \KsK)=(1.450\pm0.035)\%$ and $\mathcal{B}(\Ds \to \KKpi)=(5.37\pm0.10)\%$ are the branching fractions for $\Ds$ decaying to $\KsK$ and $\KKpi$, respectively; $\mathcal{B}(\Ks \to \pipi)=(69.20 \pm 0.05)\%$ represents the branching fraction of $\Ks$ to $\pipi$~\cite{PDG}. Figure~\ref{Fig:4680_bestfit}
shows the fit results with $\Gamma_{X}$ set to 50~$\mev$ at $\sqs=4.682~\gev$. The obtained signal yields are listed in Table~\ref{table:up_final} for each data sample, along with the corresponding statistical significance determined by variations in likelihoods and degrees of freedom with and without the signal.

Since the largest significance of the signal among all c.m.\ energies is $3.5\sigma$ at $\sqrt{s}=4.682$ GeV, we determine the upper limits of the number of signal events $N^{UL}_{KK\pi}$ at each energy point. Utilizing the Bayesian method~\cite{Bayesian}, the likelihood distribution $L(x)$ is determined by repeating the fit changing the number of expected signal events. The upper limits at $90 \%$ confidence level (C.L.) are determined by the equation $\int_0^{N_{KK\pi}^{\mathrm{UL}}} L(x) d x / \int_0^{\infty} L(x) d x=0.9$. The upper limits of the Born cross section for $\EE \to \g X$ multiplied by the branching fraction of $X \to \Ds \Dsone$ is calculated by 
\begin{equation}\label{eq:cal_up}
\sigma^{UL} \cdot \mathcal{B}= \frac{N_{KK\pi}^{UL}}{\mathcal{L}_{int} \cdot(1+\delta) \cdot \frac{1}{|1-\Pi|^2} \cdot \epsilon_{KK\pi}   \cdot \mathcal{B}_{1} \cdot \mathcal{B}_{2}},
\end{equation}
where $\mathcal{L}_{int}$ is the integrated luminosity at each energy point, $1+\delta$ is the ISR correction factor based on the QED calculation with 1\% accuracy~\cite{radi_corre} and obtained from the KKMC generator, $\frac{1}{|1-\Pi|^2}$ is the vacuum polarization factor (VP) from Ref.~\cite{VP}, $\mathcal{B}_{1}=(5.37\pm0.10)\%$ \cite{PDG} and $\mathcal{B}_{2}=(35.9\pm5.9)\%$ \cite{ref:whp} are the branching fractions for the $\Ds \to \KKpi$ and $\Dsone \to \Dstar \KM$ decays.

The upper limits of the product of the cross section and branching fraction $\sigma({\EE \to \g X}) \cdot \mathcal{B}(X \to \Ds \Dsone +c.c.)$ determined at 90\% C.L., as well as the other input values, are shown in Table \ref{table:up_final}, where only the statistical uncertainty is considered.

\section{systematic uncertainties}
\label{sec_sys}
The sources of systematic uncertainties in the determination of the upper limits of the Born cross-sections $\sigma^{UL} \cdot \mathcal{B}$ are classified into two categories: the multiplicative terms and the additive terms.  
\begin{table*}[!htbp]
\setlength{\abovecaptionskip}{1pt} 
\setlength{\belowcaptionskip}{4pt}
\caption{The multiplicative systematic uncertainties associated with $KK\pi$ mode, expressed as percentages.}
\centering
%\footnotesize
\setlength{\tabcolsep}{4pt}
\renewcommand{\arraystretch}{1.2}
\begin{tabular}{c c c c c c c c c c c c c}
\hline \hline
$\sqs~(\gev)$  &4.612  &4.628  &4.641  &4.661  &4.682  &4.699  &4.74  &4.75  &4.781  &4.843  &4.918  &4.951\\
\hline
Tracking  &4.0  &4.0  &4.0  &4.0  &4.0  &4.0  &4.0  &4.0  &4.0  &4.0  &4.0  &4.0\\
PID  &4.0  &4.0  &4.0  &4.0  &4.0  &4.0  &4.0  &4.0  &4.0  &4.0  &4.0  &4.0\\
$\g$ reconstruction  &1.0  &1.0  &1.0  &1.0  &1.0  &1.0  &1.0  &1.0  &1.0  &1.0  &1.0  &1.0\\
$M_{\Kpi}$ range  &1.2  &1.2  &1.2  &1.2  &1.2  &1.2  &1.2  &1.2  &1.2  &1.2  &1.2  &1.2\\
%$M_{\KK}$ range  &0.0  &0.0  &0.0  &0.0  &0.0  &0.0  &0.0  &0.0  &0.0  &0.0  &0.0  &0.0\\
$M_{\KKpi}$ range  &0.2  &0.2  &0.2  &0.2  &0.2  &0.2  &0.2  &0.2  &0.2  &0.2  &0.2  &0.2\\
$RM_{\gDsK}$ range  &0.7  &0.7  &0.7  &0.7  &0.7  &0.7  &0.7  &0.7  &0.7  &0.7  &0.7  &0.7\\
VP  &0.1  &0.1  &0.1  &0.1  &0.1  &0.1  &0.1  &0.1  &0.1  &0.1  &0.1  &0.1\\
Luminosity  &1.0  &1.0  &1.0  &1.0  &1.0  &1.0  &1.0  &1.0  &1.0  &1.0  &1.0  &1.0\\
$\mathcal{B}(\DstoKKpi)$  &1.9  &1.9  &1.9  &1.9  &1.9  &1.9  &1.9  &1.9  &1.9  &1.9  &1.9  &1.9\\
$\mathcal{B}(\Dsone \to \Dstar \KM)$  &16.4  &16.4  &16.4  &16.4  &16.4  &16.4  &16.4  &16.4  &16.4  &16.4  &16.4  &16.4\\
Physical model  &3.6  &3.6  &3.6  &3.6  &3.6  &3.6  &3.6  &3.6  &3.6  &3.6  &3.6  &3.6\\
Kinematic fit  &5.8  &6.2  &8.1  &4.7  &5.5  &3.4  &3.0  &3.1  &3.1  &7.6  &6.7  &6.9\\
$(1+\delta)\cdot \epsilon_{KK\pi}$  &0.5  &0.4  &0.8  &3.0  &3.3  &5.5  &4.8  &6.0  &6.9  &4.3  &4.2  &4.9\\
\hline
Total  &18.8  &19.0  &19.7  &18.8  &19.0  &19.1  &18.8  &19.2  &19.5  &19.9  &19.6  &19.8\\
\hline \hline
\end{tabular}
\label{table:multi_uncer}
\end{table*}

\begin{table*}[!htbp]
\setlength{\abovecaptionskip}{1pt} 
\setlength{\belowcaptionskip}{4pt}
\caption{The uncertainties of the correlation coefficient $\Delta$($f_{sig}$)  evaluated at each energy point, expressed as percentages.}
\centering
%\footnotesize
\setlength{\tabcolsep}{4pt}
\renewcommand{\arraystretch}{1.2}
\begin{tabular}{c c c c c c c c c c c c c}
\hline \hline
$\sqs~(\gev)$  &4.612  &4.628  &4.641  &4.661  &4.682  &4.699  &4.740  &4.750  &4.781  &4.843  &4.918  &4.951\\
\hline
tracking  &2.0  &2.0  &2.0  &2.0  &2.0  &2.0  &2.0  &2.0  &2.0  &2.0  &2.0  &2.0\\
PID  &2.0  &2.0  &2.0  &2.0  &2.0  &2.0  &2.0  &2.0  &2.0  &2.0  &2.0  &2.0\\
$\Ks$ reconstruction  &2.0  &2.0  &2.0  &2.0  &2.0  &2.0  &2.0  &2.0  &2.0  &2.0  &2.0  &2.0\\
Physical model  &4.8  &4.8  &4.8  &4.8  &4.8  &4.8  &4.8  &4.8  &4.8  &4.8  &4.8  &4.8\\
$M_{\Kpi}$ range  &1.2  &1.2  &1.2  &1.2  &1.2  &1.2  &1.2  &1.2  &1.2  &1.2  &1.2  &1.2\\
%$M_{\KK}$ range  &0.0  &0.0  &0.0  &0.0  &0.0  &0.0  &0.0  &0.0  &0.0  &0.0  &0.0  &0.0\\
$M_{\pipi}$ range  &0.3  &0.3  &0.3  &0.3  &0.3  &0.3  &0.3  &0.3  &0.3  &0.3  &0.3  &0.3\\
$M_{\KKpi/\KsK}$ range  &0.6  &0.6  &0.6  &0.6  &0.6  &0.6  &0.6  &0.6  &0.6  &0.6  &0.6  &0.6\\
$RM_{\gDsK}$ range  &0.7  &0.7  &0.7  &0.7  &0.7  &0.7  &0.7  &0.7  &0.7  &0.7  &0.7  &0.7\\
Kinematic fit  &5.5  &5.8  &7.6  &4.7  &5.2  &3.3  &2.9  &3.1  &3.0  &8.4  &8.2  &6.5\\
$\mathcal{B}(\DstoKsK)$  &2.4  &2.4  &2.4  &2.4  &2.4  &2.4  &2.4  &2.4  &2.4  &2.4  &2.4  &2.4\\
$\mathcal{B}(\Ks\to \pipi)$  &0.1  &0.1  &0.1  &0.1  &0.1  &0.1  &0.1  &0.1  &0.1  &0.1  &0.1  &0.1\\
$\mathcal{B}(\DstoKKpi)$  &1.9  &1.9  &1.9  &1.9  &1.9  &1.9  &1.9  &1.9  &1.9  &1.9  &1.9  &1.9\\
\hline
Total  &8.8  &9.0  &10.2  &8.3  &8.6  &7.6  &7.5  &7.5  &7.5  &10.8  &10.7  &9.5\\
\hline \hline
\end{tabular}
\label{table:uncer_eff_ratio}
\end{table*}
\subsection{Multiplicative systematic uncertainties}\label{sec:sys_multi}
The multiplicative uncertainties include tracking, PID, $\Ks$ reconstruction, photon reconstruction, range requirements for $M_{\Kpi}$, $M_{\KK}$, $M_{\pipi}$, $M_{\KKpi/\KsK}$ and $RM_{\gDsK}$, VP factor, luminosity, branching fractions of intermediate states, physical model, kinematic fit, and $(1+\delta)\cdot \epsilon_{KK\pi}$.

The systematic uncertainty for tracking is assigned to be $1.0 \%$ for each $\pi/ K$ track ~\cite{Uncer_track_PID_Ks}, while the uncertainty for PID is assigned to be $1.0 \%$~\cite{Uncer_track_PID_Ks}. 

We quote $2.0 \%$ as the systematic uncertainty caused by $\Ks$ reconstruction, based on Ref.~\cite{Uncer_track_PID_Ks}, in which the same $\Ks$ reconstruction method is applied. 

The uncertainty due to photon detection efficiency is determined to be 1.0\% per photon~\cite{ref:gDstar}, according to a study using as control sample $J/\psi \to \rho\pi$ events.

The systematic uncertainties due to the range requirements on  $M_{\Kpi}$, $M_{\KK}$, and $M_{\KKpi}$ for the $KK\pi$ mode, and on $M_{\pipi}$ and $M_{\KsK}$ for the $K_SK$ mode, stem from the resolution difference between data and MC and  are determined by comparing the signal shapes of data and MC; they are found to be $1.2\%$, $<0.1\%$ (negligible), $0.2\%$, $0.3\%$, and $0.6\%$, respectively. Since a simultaneous fit is performed in this analysis, only the uncertainties associated with $M_{\Kpi}$, $M_{\KK}$, and $M_{\KKpi}$ contribute to the global multiplicative uncertainty, even though all of them affect the correlation coefficient $f_{sig}$.
Since no significant signal is observed in the recoil mass of $\gDsK$ in data, a different method is employed to study the systematic uncertainty from the $RM_{\gDsK}$ mass window requirement; we smear the $\g$ energy distribution  by a Gaussian with 1\% uncertainty and reconstruct the smeared recoil mass $RM^{smear}_{\gDsK}$. The difference in efficiency by applying the same requirement in the $RM_{\gDsK}$ and $RM^{smear}_{\gDsK}$ distributions is taken as the related systematic uncertainty. We assign 0.7\% and 0.2\% for the range requirement of $RM_{\gDsK}$ in $KK\pi$ mode and $\Ks K$ mode, respectively.

The uncertainty from the vacuum polarization factor is less than 0.1\%~\cite{VP}.

The integrated luminosity is measured using Bhabha events, with an uncertainty of about 1.0\% at each energy point~\cite{XYZ_data}. 

The systematic uncertainties associated with the branching fractions $\mathcal{B}(\DstoKKpi)$, $\mathcal{B}(\DstoKsK)$, $\mathcal{B}(\Ks \to \pipi)$, and $\mathcal{B}(\Dsone \to \Dstar \KM)$ are quoted as 1.9\%~\cite{PDG}, 2.4\%~\cite{PDG}, 0.1\%~\cite{PDG}, and 16.4\%~\cite{ref:whp}, respectively.

The uncertainty associated with the physical model of the $\EE \to \g X$ process is estimated by changing the PHSP to a helicity amplitude (HELAMP) model~\cite{ref:evtgen} when generating the signal MC sample. The differences in efficiency between the PHSP and the HELAMP model, 3.6\% for the $KK\pi$ mode and 3.1\% for the $\Ks K$ mode, are considered as systematic uncertainties. The PHSP model, which results with a lower efficiency, is adopted as the nominal model for a conservative measurement.

To study the systematic uncertainty caused by the kinematic fit, we correct the helix parameters of charged tracks in the MC simulation \cite{helix_correct}. The efficiency difference with and without the helix correction, as shown in Table \ref{table:multi_uncer}, is taken as the systematic uncertainty.

\begin{table*}[!htbp]
\setlength{\abovecaptionskip}{1pt} 
\setlength{\belowcaptionskip}{4pt}
\caption{The results of $\sigma({\EE \to \g X}) \cdot \mathcal{B}(X \to \Ds \Dsone +c.c.)$ under different assumptions of $X$'s width ($\Gamma_{X}=25, 50, 75, 100$~$\mev$) including the systematic uncertainties. The $N_{KK\pi}^{UL'}$ and $\sigma^{UL}_{sys} \cdot \mathcal{B}$ values represent the upper limits of the number of signal events and of the product of the cross-section and branching fraction  at 90\% C.L.}
\centering
%\footnotesize
\setlength{\tabcolsep}{2pt}
\renewcommand{\arraystretch}{1.2}
\begin{tabular}{c c c c c c c c c}
\hline\hline
 $\Gamma_{X}$  & \multicolumn{2}{c}{25~$\mev$}  & \multicolumn{2}{c}{50~ $\mev$} &\multicolumn{2}{c}{75~$\mev$} &\multicolumn{2}{c}{100~$\mev$}\\
\hline
$\sqs~(\gev)$ &$N_{KK\pi}^{UL'}$   &$\sigma^{UL}_{sys} \cdot \mathcal{B}$ (pb)    &$N_{KK\pi}^{UL'}$   &$\sigma^{UL}_{sys} \cdot \mathcal{B}$ (pb)  &$N_{KK\pi}^{UL'}$   &$\sigma^{UL}_{sys} \cdot \mathcal{B}$ (pb)   &$N_{KK\pi}^{UL'}$   &$\sigma^{UL}_{sys} \cdot \mathcal{B}$ (pb) \\
\hline
4.612  &1.7  &26.9  &1.7  &26.9  &1.7  &26.9  &1.7  &26.9  \\
4.628  &2.8  &8.6  &2.8  &8.6  &3.1  &9.5  &3.1  &9.5  \\
4.641  &4.3  &12.8  &5.5  &16.3  &5.9  &17.5  &5.8  &17.2  \\
4.661  &2.7  &12.7  &2.8  &13.1  &3.3  &15.4  &3.2  &15.0  \\
4.682  &10.2  &10.7  &13.1  &13.8  &14.7  &15.5  &15.4  &16.2  \\
4.699  &4.8  &12.9  &5.7  &15.3  &6.3  &16.9  &6.7  &18.0  \\
4.740  &1.8  &14.2  &1.9  &15.0  &1.9  &15.0  &1.9  &15.0  \\
4.750  &3.4  &12.0  &4.4  &15.4  &4.7  &16.5  &4.8  &16.8  \\
4.781  &1.9  &4.6  &2.1  &5.1  &2.4  &5.8  &2.5  &6.1  \\
4.843  &1.8  &4.6  &1.8  &4.6  &1.9  &4.8  &2.0  &5.1  \\
4.918  &5.2  &30.6  &5.3  &31.1  &6.8  &39.9  &7.0  &41.0  \\
4.951  &1.7  &13.5  &1.7  &13.5  &1.7  &13.5  &1.7  &13.5  \\
\hline \hline
\end{tabular}
\label{comp_cross}
\end{table*}

In the nominal results, the detection efficiencies and ISR corrections $(1+\delta)\cdot \epsilon_{KK\pi}$ are obtained from MC sample, with the cross-section line shape described by $E_{\g}^{3}/s$. The uncertainty from the input cross-section line shape is estimated by using the Breit-Wigner function of the $Y(4660)$ state instead. The changes in the resultant Born cross-sections, shown in Table \ref{table:multi_uncer}, are taken as the systematic uncertainty from the $(1+\delta)\cdot \epsilon_{KK\pi}$ term.

Based on Eq.~\ref{eq:cal_up}, all the multiplicative systematic uncertainties associated with the $KK\pi$ mode for each energy point are listed in Table~\ref{table:multi_uncer}, and the total systematic uncertainties are obtained by adding them in quadrature assuming no correlation among all the sources. To take into account the multiplicative systematic uncertainties in the upper limit calculation, the normalized likelihood distributions are smeared with a Gaussian function with a mean of
$\epsilon_{KK\pi}$ and a standard deviation of $\delta^{abs}_{\epsilon}=\epsilon_{KK\pi} \cdot \delta_{\epsilon}^{rel}$, where $\delta^{abs}_{\epsilon}$ and $\delta_{\epsilon}^{rel}$ are the absolute and relative multiplicative systematic uncertainties, respectively. The smeared likelihood is defined as
\begin{equation}
\mathrm{L}(N_{KK\pi})=\int_0^1 \mathrm{~L}^{\prime}(\frac{\epsilon}{\epsilon_{KK\pi}} N_{KK\pi}) \frac{1}{\sqrt{2 \pi} \delta_{\epsilon}^{abs}} e^{-\frac{(\epsilon-\epsilon_{KK\pi})^2}{2 \delta_{\epsilon}^{abs}}} d \epsilon.
\end{equation}

\subsection{Additive systematic uncertainties}\label{section:additive}
The additive systematic uncertainties arise from the fit process and mainly include the uncertainties associated with the background shape and the correlation coefficient $f_{sig}$, which is related to the efficiency ratio $\epsilon_{\Ks K}/\epsilon_{KK\pi}$, as well as the branching fractions $\mathcal{B}(\DstoKsK)$, $\mathcal{B}(\Ks \to \pipi)$, and $\mathcal{B}(\DstoKKpi)$. 

We estimate the systematic uncertainty from background shape by using a first-order Chebyshev function instead of the second-order Chebyshev function in the $RM_{\g\Ds}$ projection, and a second-order Chebyshev function instead of the anti-Argus function~\cite{Argus} in the $RM_{\g}$ projection to describe the background shapes. The highest upper limit of the variations in the background shapes is assigned as the associated systematic uncertainty.

We account for the systematic uncertainty arising from the efficiency ratio $\epsilon_{\Ks K}/\epsilon_{KK\pi}$ and the branching fractions $\mathcal{B}(\DstoKsK)$, $\mathcal{B}(\Ks \to \pipi)$, and $\mathcal{B}(\DstoKKpi)$ in the simultaneous fit, employing the same method as the multiplicative systematic uncertainties in Section \ref{sec:sys_multi}. Note that the uncertainties associated with tracking and PID, which contribute to the efficiency ratio $\epsilon_{\Ks K}/\epsilon_{KK\pi}$, are partially correlated; therefore, their effects are partially canceled. The uncertainties related to the $\g$ reconstruction are fully canceled out. Other uncertainty sources of the efficiency ratio $\epsilon_{\Ks K}/\epsilon_{KK\pi}$ associated with the physical model, mass interval requirement, and kinematic fit are considered as independent in the two modes. All the uncertainties from different sources are treated as independent, and the total systematic uncertainty of $f_{sig}$ is obtained by summing them in quadrature. The systematic uncertainties from $f_{sig}$ at each energy point are presented in Table~\ref{table:uncer_eff_ratio}. We incorporate the $f_{sig}$ and its related uncertainty at each energy point in the simultaneous fit, adopting the highest upper limit as the value considering the associated systematic uncertainty.

\begin{figure}[!htbp]
\centering   
\includegraphics[angle=0,width=0.45\textwidth]{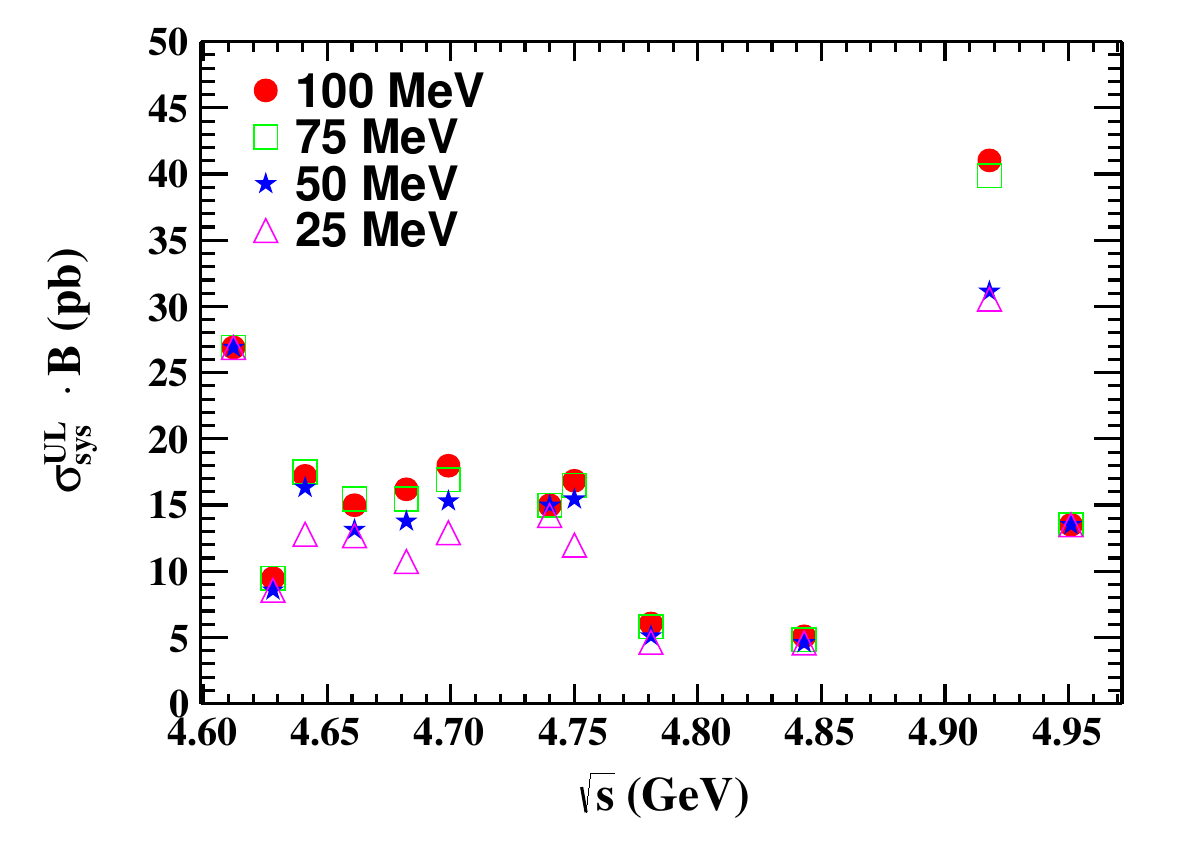}
\caption{The upper limits of the product of the cross section and branching fraction $\sigma(\EE \to \g X) \cdot \mathcal{B}(X \to \Ds \Dsone +c.c.)$ at each energy points assuming the width  $\Gamma_{X}=25$, $50$, $75$, or $100$~$\mev$ at 90\% C.L..}
\label{plot_cross_line}
\end{figure}

After considering the systematic uncertainties, Table~\ref{comp_cross} and Fig.~\ref{plot_cross_line} represent the upper limits of the product of the cross-section and the branching fraction $\sigma({\EE \to \g X}) \cdot \mathcal{B}(X \to \Ds \Dsone)$, assuming $\Gamma_{X}$ is $25~\mev$, $50~\mev$, $75~\mev$, or $100~\mev$.
\section{Summary}
Based on $\EE$ collision data at c.m.\ energies from 4.612 to 4.951~$\gev$ collected by the BESIII spectrometer, corresponding to an integrated luminosity of $5.8~\mathrm{fb^{-1}}$, the exotic molecular state $X$ ($\Jpc=1^{-+}$) is searched for the first time via the process $\EE \to \g X, X \to \Ds \Dsone$.
No significant signal is observed and the upper limits of $\sigma(\EE \to \g X) \cdot \mathcal{B}(X \to \Ds \Dsone)$ are determined at 90\% C.L., assuming $M_{X}=4.503~\gevcs$ and $\Gamma_{X}= 25$, $50$, $75$, or $100$~$\mev$.
The obtained upper limits are shown in Table~\ref{comp_cross} and Fig.~\ref{plot_cross_line}. These upper limits range from $5$~pb to $45$~pb, while typically $15$~pb around $4.68~\gev$. Assuming the total cross section of $e^+ e^- \to Y(4660)$ is about $300$~pb~\cite{PDG} and the branching faction of $Y(4660) \to \gamma X$ being $0.1\,\%$ (same to $\mathcal{B}(Y(4260) \to \gamma X')$ estimated in Ref.~\cite{WQ}, where $X'$ decays into non-strange D meson pairs), the product of the cross-section and the branching fraction is $0.3$~pb, which is 50 times lower than the upper limit determined in this analysis.  At $\sqrt{s}=4.682$~GeV, there is evidence of a signal with $3.5\sigma$ significance when $\Gamma_{X}= 50~\mev$. However, this evidence may only be a local and statistical fluctuation. Therefore, more experimental data are needed to confirm or exclude it by considering the global significance and the systematic uncertainties.

\begin{acknowledgments}
The BESIII Collaboration thanks the staff of BEPCII and the IHEP computing center for their strong support. This work is supported in part by National Key R\&D Program of China under Contracts Nos. 2020YFA0406300, 2020YFA0406400, 2023YFA1606000; National Natural Science Foundation of China (NSFC) under Contracts Nos. 12275058, 11635010, 11735014, 11935015, 11935016, 11935018, 12025502, 12035009, 12035013, 12061131003, 12192260, 12192261, 12192262, 12192263, 12192264, 12192265, 12221005, 12225509, 12235017, 12361141819; the Chinese Academy of Sciences (CAS) Large-Scale Scientific Facility Program; the CAS Center for Excellence in Particle Physics (CCEPP); Joint Large-Scale Scientific Facility Funds of the NSFC and CAS under Contract No. U1832207; 100 Talents Program of CAS; The Institute of Nuclear and Particle Physics (INPAC) and Shanghai Key Laboratory for Particle Physics and Cosmology; German Research Foundation DFG under Contracts Nos. FOR5327, GRK 2149; Istituto Nazionale di Fisica Nucleare, Italy; Knut and Alice Wallenberg Foundation under Contracts Nos. 2021.0174, 2021.0299; Ministry of Development of Turkey under Contract No. DPT2006K-120470; National Research Foundation of Korea under Contract No. NRF-2022R1A2C1092335; National Science and Technology fund of Mongolia; National Science Research and Innovation Fund (NSRF) via the Program Management Unit for Human Resources \& Institutional Development, Research and Innovation of Thailand under Contracts Nos. B16F640076, B50G670107; Polish National Science Centre under Contract No. 2019/35/O/ST2/02907; Swedish Research Council under Contract No. 2019.04595; The Swedish Foundation for International Cooperation in Research and Higher Education under Contract No. CH2018-7756; U. S. Department of Energy under Contract No. DE-FG02-05ER41374
\end{acknowledgments}

%\nocite{*}
\bibliography{draft}

\end{document}